# Finite-State Dimension and Lossy Decompressors

David Doty* Philippe Moser†

**Abstract**

This paper examines information-theoretic questions regarding the difficulty of compressing data versus the difficulty of decompressing data and the role that information loss plays in this interaction. Finite-state compression and decompression are shown to be of equivalent difficulty, even when the decompressors are allowed to be lossy.

Inspired by Kolmogorov complexity, this paper defines the optimal *decompression* ratio achievable on an infinite sequence by finite-state decompressors (that is, finite-state transducers *outputting* the sequence in question). It is shown that the optimal compression ratio achievable on a sequence $S$ by any *information lossless* finite state compressor, known as the finite-state dimension of $S$, is equal to the optimal decompression ratio achievable on $S$ by *any* finite-state decompressor. This result implies a new decompression characterization of finite-state dimension in terms of *lossy* finite-state transducers.

## 1 Introduction

This paper addresses the fundamental information-theoretic question: is the problem of compressing data to a short representation of the same difficulty as the problem of decompressing data from a short representation? It is known that for certain cases admitting sufficient computational resources, both problems are indeed of equivalent difficulty. For example, consider the case of polynomial-space-bounded Kolmogorov complexity [11]. The shortest program computing a string $x$ in polynomial space can be computed from $x$ in polynomial space, by reusing space to conduct an exponential time search for short polynomial space programs for $x$. However, this result is not known to hold at lower levels of complexity, such as polynomial-time-bounded Kolmogorov complexity. At the level of unbounded computation, there is a known incongruity between compression and

*Department of Computer Science, Iowa State University, Ames, IA 50011 USA. ddoty *at* iastate *dot* edu.

†Dept de Informática e Ingeniería de Sistemas, Centro Politécnico Superior, Zaragoza, Spain. mosersan *at* gmail *dot* com. This work was partially supported by subvenciones para grupos de investigación Gobierno de Aragón UZ-T27 and subvenciones de fomento de movilidad Gobierno de Aragón MI31/2005.

decompression: a string is computable from its shortest program, but the converse does not hold.

This paper settles the case at a level of computational complexity lower even than polynomial time: the finite-state level. It was already known [9, 10] that, if attention is restricted to *information lossless (IL)* finite-state transducers [9], compression and decompression are of equivalent difficulty. Our main result shows that we need not restrict attention to IL transducers to obtain this equivalence. Inspired by Kolmogorov complexity [11], we define the optimal *decompression* ratio achievable on an infinite sequence by (possibly lossy) finite-state transducers acting as *decompressors*. Our result implies that this quantity is equal to the optimal compression ratio achievable on the sequence with IL finite-state compressors.

More precisely, given an infinite sequence $S$, Ziv and Lempel [20] defined the *finite-state strong dimension* of $S$ (called the *finite-state compressibility* of $S$ in [20]) to be

$$\mathrm{Dim}_{\mathrm{FS}}(S) = \lim_{k\to\infty} \limsup_{n\to\infty} \frac{\mathrm{C}^k_{\text{ILFS-LZ}}(S \upharpoonright n)}{n}, \tag{1.1}$$

where $\mathrm{C}^k_{\text{ILFS-LZ}}(S \upharpoonright n)$ is the length of smallest string output by any information lossless finite-state transducer (ILFST) with at most $k$ states, when given $S \upharpoonright n$ as input. An analogous quantity, the *finite-state dimension* $\dim_{\mathrm{FS}}(S)$ of $S$ [3, 1], is defined similarly, by replacing the limit superior in (1.1) with a limit inferior. Finite-state dimension and strong dimension are so called because they have been shown [3, 1] to be finite-state effectivizations, respectively, of classical Hausdorff dimension [5] and packing dimension [18, 17], the two most widely-used fractal dimensions. Each admits a host of different characterizations, in terms of finite-state gamblers [3, 1], entropy rates [20, 2], information lossless finite-state compressors [20, 3, 1], and finite-state log-loss predictors [7]. This indicates that finite-state dimension is a robust and stable quantity that truly measures the information density of a sequence as perceived by finite-state machines, to a certain extent independent of the details of the particular finite-state machine model under consideration.

An ILFST is a finite-state transducer (FST) that must create an output from which the input can be uniquely recovered, whereas a general FST has no such restriction. An ILFST $T$ therefore cannot output small strings on most inputs, which limits which strings $T$ can significantly compress. By contrast, the quantity $\mathrm{C}^k_{\mathrm{FS}}(x)$, defined similarly to $\mathrm{C}^k_{\text{ILFS-LZ}}(x)$, but without the IL requirement, is trivially equal to 0 for all strings $x$, because a 1-state FST that always outputs the empty string compresses every string to length 0. This FST "cheats" by throwing away information contained in its input. Requiring the FST to be IL prevents this cheating and limits the compression performance of the FST. From this perspective, we consider the following two questions.

1. Does the characterization of finite-state dimension still hold if we consider decompressors instead of compressors? That is, suppose the FST, rather than aiming to compress the given sequence to a more compact sequence, is instead aiming to expand a compact sequence into the given sequence.

2. If the answer to question 1 is yes, is it mandatory that the decompressor be IL in order to characterize finite-state dimension? In other words, would allowing a decompressor to be an arbitrary FST afford it more power to decompress than if it were IL, as in the case of compression?

An affirmative answer to question 1 follows in a straightforward manner from the well-known result [9, 10] that every ILFST computes a function whose inverse is computable by another ILFST (in a technical sense described in Theorem 3.3). The answer to question 2 is less obvious. There are clearly functions that are computable by a FST but not computable by any ILFST (for example, any constant function). Informally, question 2 asks, can a FST acting as a decompressor improve its performance – i.e. output a larger string than otherwise possible – by throwing away information?

The main result of this paper answers question 2 negatively. We show that given a lossy FST $T$, there is an ILFST $T'$ with the property that, for all strings $x$, the shortest input to $T'$ that outputs $x$ is no larger than the shortest input to $T$ that outputs $x$. Therefore, while $T'$ cannot do everything that $T$ can do, it can decompress as effectively as $T$. The intuitive reason this is possible is that, although $T$ is lossy, *optimally compressed* input to $T$ follows an "information lossless path" through $T$. We construct $T'$ to preserve such IL paths, while amending only the "lossy paths" through $T$ in order to make it IL.

This result implies that the finite-state dimension of a sequence can be characterized in terms of the optimal decompression ratio achieved on the sequence by *any* finite-state decompressor. More precisely, define $\mathrm{D}^k_{\mathrm{FS}}(x)$ to be the length of the smallest string that produces $x$ as output, when given as input to some FST that requires at most $k$ bits to describe in a standard binary representation of FST's.[1] We show that the finite-state strong dimension of a sequence $S$ can be characterized by replacing $\mathrm{C}^k_{\text{ILFS-LZ}}$ with $\mathrm{D}^k_{\mathrm{FS}}$ in (1.1) (and analogously for finite-state dimension).

One interpretation of $\mathrm{D}^k_{\mathrm{FS}}$ is as a finite-state adaptation of Kolmogorov complexity, with Kolmogorov complexity considered to measure "optimal decompression" at the level of unbounded computation. From this perspective, our finite-state dimension characterization mirrors previously known characterizations of other effective dimensions, such as constructive dimension [13, 14], computable dimension, and various space-bounded dimensions such as polynomial-space dimension [12, 6], in terms of Kolmogorov complexity or space-bounded Kolmogorov complexity [11]. It remains an open question whether polynomial-time dimension [12] can be characterized in terms of polynomial-time Kolmogorov complexity (see [8] for a summary of recent progress on this question).

After writing this paper, the authors became aware of a very similar result proven by Lempel, Sheinwald, and Ziv [16]. Therefore, our proof of Theorem 3.11 may be considered a new proof of Corollary 2.3 of [16].

[1] Unlike in the quantity $\mathrm{C}^k_{\text{ILFS-LZ}}$ used by Ziv and Lempel, the $k$ in $\mathrm{D}^k_{\mathrm{FS}}$ does not represent the number of states of the FST, but rather its total description length. This discrepancy is explained in §2.

# 2 Preliminaries

## 2.1 Notation

Throughout this paper, $\Sigma$ is a finite *alphabet.* $\mathbb{N}$ is the set of all nonnegative integers. All *strings* are elements of $\Sigma^*$, and all *sequences* are elements of $\Sigma^\infty$. For all $x \in \Sigma^*$, we write $|x|$ to denote the *length* of $x$. For all $k \in \mathbb{N}$, $\Sigma^k$, $\Sigma^{\leq k}$, and $\Sigma^{<k}$ are the set of strings of length exactly $k$, at most $k$, and less than $k$, respectively. $\lambda$ denotes the empty string. If $x$ is a string or sequence and $i, j$ are integers, $x[i\,..\,j]$ denotes the string consisting of the $i^{\text{th}}$ through $j^{\text{th}}$ symbols in $x$, with $x[i\,..\,j] = \lambda$ if $j < i$, noting that $x[0]$ is the leftmost symbol in $x$, and we write $x \upharpoonright n$ to denote $x[0\,..\,n-1]$. If $w$ is a string and $x$ is a string or sequence, we say $w$ is a *prefix* of $x$, and we write $w \sqsubseteq x$, if $x = wu$ for some $u \in \Sigma^*$, and we write $w \sqsubset x$ if $w \sqsubseteq x$ and $w \neq x$. We say $w$ is a *suffix* of $x$ if $x = uw$ for some $u \in \Sigma^*$, and we say $w$ is a *proper suffix* of $x$ if $w$ is a suffix of $x$ and $w \neq x$. For a set $X \subseteq \Sigma^*$, we say $X$ is *suffix-free* if, for all $x, y \in X$, $x$ is not a proper suffix of $y$.

## 2.2 Finite-State Compression

In this section, we develop a notion of finite-state compression and decompression that serves to measure the optimal amount by which strings and sequences can be compressed and decompressed by finite-state transducers. We base our model of finite-state transducers on that studied in [3], which was introduced in a similar form by Shannon [15] and investigated by Huffman [9] and Ziv and Lempel [19]. Kohavi [10] gives an extensive treatment of the subject.

A *finite-state transducer (FST)* is a 4-tuple

$$T = (Q, \delta, \nu, q_0),$$

where

- $Q$ is a nonempty, finite set of *states*,
- $\delta : Q \times \Sigma \to Q$ is the *transition function*,
- $\nu : Q \times \Sigma \to \Sigma^*$ is the *output function*,
- $q_0 \in Q$ is the *initial state*.

Furthermore, we assume that every state in $Q$ is reachable from $q_0$. Given $q_1, q_2 \in Q$ and $a \in \Sigma$ such that $\delta(q_1, a) = q_2$, we refer to the triple $(q_1, a, q_2)$ as a *transition arrow* in the directed graph representing the FST, in order to emphasize where the arrow starts and ends, and what input symbol causes the FST to follow it. By this interpretation, if $a \neq a'$ but $q_2 = \delta(q_1, a) = \delta(q_1, a')$, then $(q_1, a, q_2)$ and $(q_1, a', q_2)$ constitute different transition arrows, even though they start and end at the same states.

For all $x \in \Sigma^*$ and $a \in \Sigma$, define the *extended transition function* $\widehat{\delta} : \Sigma^* \to Q$ by the recursion

$$\begin{aligned} \widehat{\delta}(\lambda) &= q_0, \\ \widehat{\delta}(xa) &= \delta(\widehat{\delta}(x), a). \end{aligned}$$

For $x \in \Sigma^*$, we define the *output* of $T$ on $x$ to be the string $T(x)$ defined by the recursion

$$\begin{aligned} T(\lambda) &= \lambda, \\ T(xa) &= T(x)\nu(\widehat{\delta}(x), a) \end{aligned}$$

for all $x \in \Sigma^*$ and $a \in \Sigma$. Given any FST $T$, we say $\pi \in \Sigma^*$ is a *minimal program* for $T$ if, for all $\pi' \in \Sigma^{<|\pi|}$, $T(\pi) \neq T(\pi')$; i.e., $\pi$ is a shortest input to $T$ that produces the output $T(\pi)$.

A FST can trivially act as an "optimal compressor" by outputting $\lambda$ on every transition arrow, but this is, of course, a useless compressor, because the input cannot be recovered. A FST $T = (Q, \delta, \nu, q_0)$ is *information lossless (IL)* if the function $x \mapsto (T(x), \widehat{\delta}(x))$ is one-to-one; i.e., if the output and final state of $T$ on input $x$ uniquely identify $x$. An *information lossless finite-state transducer (ILFST)* is a FST that is IL. We write FST to denote the set of all finite-state transducers, and we write ILFST to denote the set of all information lossless finite-state transducers.

Let $S \in \Sigma^\infty$. The *finite-state dimension* [3] and the *finite-state strong dimension* [1] of $S$ are respectively defined

$$\dim_{\mathrm{FS}}(S) = \inf_{T \in \mathrm{ILFST}} \liminf_{n \to \infty} \frac{|T(S \upharpoonright n)|}{n},$$

and

$$\mathrm{Dim}_{\mathrm{FS}}(S) = \inf_{T \in \mathrm{ILFST}} \limsup_{n \to \infty} \frac{|T(S \upharpoonright n)|}{n}.$$

Intuitively, the finite-state dimension (resp. strong dimension) of a sequence represents the optimal best-case (resp. worst-case) compression ratio achievable on the sequence with any information lossless finite-state *compressor*. (This is a different definition of finite-state dimension than that given in the Introduction; Lemma 3.1 tells us that they are in fact equivalent.)

Fix some standard binary representation $\sigma_T \in \{0,1\}^*$ of each FST $T$, and define $|T| = |\sigma_T|$. For all $k \in \mathbb{N}$, define

$$\begin{aligned} \mathrm{FST}^{\leq k} &= \{\, T \in \mathrm{FST} \mid |T| \leq k \,\}, \\ \mathrm{ILFST}^{\leq k} &= \{\, T \in \mathrm{ILFST} \mid |T| \leq k \,\}, \\ \mathrm{ILFST}^{\leq k\text{-state}} &= \{\, T = (Q, \delta, \nu, q_0) \in \mathrm{ILFST} \mid |Q| \leq k \,\}. \end{aligned}$$

Note that, for all $k \in \mathbb{N}$, $\mathrm{ILFST}^{\leq k} \subseteq \mathrm{ILFST}^{\leq k\text{-state}}$ and $\mathrm{ILFST}^{\leq k} \subseteq \mathrm{FST}^{\leq k}$.

We next define quantities that may be considered parameterized finite-state analogs of Kolmogorov complexity. For all $k \in \mathbb{N}$ and $x \in \Sigma^*$, define the *k-finite-state decompression complexity* of $x$ by

$$\mathrm{D}^k_{\mathrm{FS}}(x) = \min_{\pi \in \Sigma^*} \left\{ \ |\pi| \ \middle| \ (\exists T \in \mathrm{FST}^{\leq k}) \ T(\pi) = x \ \right\},$$

the *k-IL-finite-state decompression complexity* of $x$ by

$$\mathrm{D}^k_{\mathrm{ILFS}}(x) = \min_{\pi \in \Sigma^*} \left\{ \ |\pi| \ \middle| \ (\exists T \in \mathrm{ILFST}^{\leq k}) \ T(\pi) = x \ \right\},$$

the *k-IL-finite-state compression complexity* of $x$ by

$$\mathrm{C}^k_{\mathrm{ILFS}}(x) = \min_{\pi \in \Sigma^*} \left\{ \ |\pi| \ \middle| \ (\exists T \in \mathrm{ILFST}^{\leq k}) \ T(x) = \pi \ \right\},$$

and the *k-IL-finite-state Lempel-Ziv compression complexity* [20] of $x$ by

$$\mathrm{C}^k_{\text{ILFS-LZ}}(x) = \min_{\pi \in \Sigma^*} \left\{ \ |\pi| \ \middle| \ (\exists T \in \mathrm{ILFST}^{\leq k\text{-state}}) \ T(x) = \pi \ \right\}.$$

# 3 Information Loss and Finite-State Decompression

The following lemma is due to Athreya, Hitchcock, Lutz, and Mayordomo [1].

**Lemma 3.1.** *For all $S \in \Sigma^\infty$,*

$$\lim_{k\to\infty} \liminf_{n\to\infty} \frac{\mathrm{C}^k_{\text{ILFS-LZ}}(S \upharpoonright n)}{n} = \inf_{T \in \mathrm{ILFST}} \liminf_{n\to\infty} \frac{|T(S \upharpoonright n)|}{n} \ (= \dim_{\mathrm{FS}}(S)),$$

*and*

$$\lim_{k\to\infty} \limsup_{n\to\infty} \frac{\mathrm{C}^k_{\text{ILFS-LZ}}(S \upharpoonright n)}{n} = \inf_{T \in \mathrm{ILFST}} \limsup_{n\to\infty} \frac{|T(S \upharpoonright n)|}{n} \ (= \mathrm{Dim}_{\mathrm{FS}}(S)).$$

For all $k \in \mathbb{N}$ and all $T \in \mathrm{ILFST}^{\leq k}$, $\mathrm{C}^k_{\text{ILFS-LZ}}(S \upharpoonright n) \leq \mathrm{C}^k_{\mathrm{ILFS}}(S \upharpoonright n) \leq |T(S \upharpoonright n)|$. By Lemma 3.1, we arrive at the following characterization of finite-state dimension.

**Observation 3.2.** *For all $S \in \Sigma^\infty$,*

$$\dim_{\mathrm{FS}}(S) = \lim_{k\to\infty} \liminf_{n\to\infty} \frac{\mathrm{C}^k_{\mathrm{ILFS}}(S \upharpoonright n)}{n},$$

*and*

$$\mathrm{Dim}_{\mathrm{FS}}(S) = \lim_{k\to\infty} \limsup_{n\to\infty} \frac{\mathrm{C}^k_{\mathrm{ILFS}}(S \upharpoonright n)}{n}.$$

We choose this characterization of finite-state dimension to investigate the relationship between compression and decompression because, in contrast to $\mathrm{C}^k_{\text{ILFS-LZ}}$, the *decompression* complexity measures $\mathrm{D}^k_{\mathrm{FS}}$ and $\mathrm{D}^k_{\mathrm{ILFS}}$ would become trivial if the transducers in $\mathrm{FST}^{\leq k}$ and $\mathrm{ILFST}^{\leq k}$ were limited only to those FST's with at most $k$ states: for each $x \in \Sigma^*$, a 1-state FST with $x$ on a transition arrow would suffice to produce $x$ from a single input symbol. Therefore, we limit the total description length of the transducer when considering $\mathrm{FST}^{\leq k}$ and $\mathrm{ILFST}^{\leq k}$, in order to account for both the number of states *and* the size of the output strings.

The following well-known theorem [9, 10] states that the function from $\Sigma^*$ to $\Sigma^*$ computed by an ILFST can be inverted – in an approximate sense – by another ILFST.

**Theorem 3.3.** *For any ILFST $T$, there exists an ILFST $T^{-1}$ and a constant $c \in \mathbb{N}$ such that, for all $x \in \Sigma^*$, $x \upharpoonright (|x| - c) \sqsubseteq T^{-1}(T(x)) \sqsubseteq x$.*

The following lemma shows that, due to Theorem 3.3, finite-state dimension can be characterized in terms of optimal *decompression* by ILFST's.

**Lemma 3.4.** *For all $S \in \Sigma^\infty$,*

$$\dim_{\mathrm{FS}}(S) = \lim_{k\to\infty} \liminf_{n\to\infty} \frac{\mathrm{D}^k_{\mathrm{ILFS}}(S \upharpoonright n)}{n},$$

*and*

$$\mathrm{Dim}_{\mathrm{FS}}(S) = \lim_{k\to\infty} \limsup_{n\to\infty} \frac{\mathrm{D}^k_{\mathrm{ILFS}}(S \upharpoonright n)}{n}.$$

*Proof.* We prove the result for $\dim_{\mathrm{FS}}$. The proof for $\mathrm{Dim}_{\mathrm{FS}}$ is analogous.

To show $\dim_{\mathrm{FS}}(S) \geq \lim\limits_{k\to\infty} \liminf\limits_{n\to\infty} \frac{\mathrm{D}^k_{\mathrm{ILFS}}(S\upharpoonright n)}{n}$, let $d > d' > \dim_{\mathrm{FS}}(S)$, and let $\epsilon = 1 - \frac{d'}{d} > 0$. By our choice of $d'$, there exists $k \in \mathbb{N}$ and $C \in \mathrm{ILFST}^{\leq k}$ such that for infinitely many $n \in \mathbb{N}$, $|C(S \upharpoonright n)| < d'n$. Let $D = C^{-1}$ and $c \in \mathbb{N}$ be given by Theorem 3.3. Thus $D \in \mathrm{ILFST}^{\leq k'}$ for some $k'$, and for every $n \in N$, $D(C(S \upharpoonright n)) = S \upharpoonright m_n$ where $n - c \leq m_n \leq n$. If $p_n = C(S \upharpoonright n)$, then for infinitely many $n \in N$, $D(p_n) = S \upharpoonright m_n$ where

$$\frac{|p_n|}{m_n} \leq \frac{|p_n|}{n-c} \leq \frac{|p_n|}{n - \epsilon n} \leq \frac{|p_n|}{n(1-\epsilon)} < \frac{d'}{1-\epsilon} = d,$$

whence $\dim_{\mathrm{FS}}(S) \geq \lim\limits_{k\to\infty} \liminf\limits_{n\to\infty} \frac{\mathrm{D}^k_{\mathrm{ILFS}}(S\upharpoonright n)}{n}$.

To show $\dim_{\mathrm{FS}}(S) \leq \lim\limits_{k\to\infty} \liminf\limits_{n\to\infty} \frac{\mathrm{D}^k_{\mathrm{ILFS}}(S\upharpoonright n)}{n}$, let $d > \lim\limits_{k\to\infty} \liminf\limits_{n\to\infty} \frac{\mathrm{D}^k_{\mathrm{ILFS}}(S\upharpoonright n)}{n}$. By choice of $d$, there exists $k \in \mathbb{N}$ and $D \in \mathrm{ILFST}^{\leq k}$ such that for any $n \in \mathbb{N}$, there exists $p_n \in \Sigma^*$ such that $D(p_n) = S \upharpoonright n$ and $|p_n| < dn$. Let $C = D^{-1}$ and $c \in \mathbb{N}$ be given by Theorem 3.3. Thus $C \in \mathrm{ILFST}^{\leq k'}$ for some $k'$, and for every $n \in N$, $C(D(p_n)) = p'_n$ where $p'_n \sqsubseteq p_n$. Hence for infinitely many $n \in \mathbb{N}$,

$$\frac{|C(S \upharpoonright n)|}{n} = \frac{|C(D(p_n))|}{n} = \frac{|p'_n|}{n} \leq \frac{|p_n|}{n} < d,$$

whence $\dim_{\text{FS}}(S) \leq \lim_{k\to\infty} \liminf_{n\to\infty} \frac{\mathrm{D}^k_{\text{ILFS}}(S\upharpoonright n)}{n}$. □

Let $T = (Q, \delta, \nu, q_0)$ be a FST. Define a *path* in $T$ to be a finite sequence $p = (p_0, a_0, p_1, a_1, \ldots, p_{n-1}, a_{n-1}, p_n)$, where $p_i \in Q$ and $a_i \in \Sigma$, satisfying, for all $0 \leq i \leq n-1$, $\delta(p_i, a_i) = p_{i+1}$. Let $|p| = n$ denote the *length* of $p$, the number of transition arrows it follows. For $0 \leq i \leq n$, define

$$p \upharpoonright i = (p_0, a_0, p_1, a_1, \ldots, p_{i-1}, a_{i-1}, p_i),$$

with $p \upharpoonright 0 = (p_0)$. Let

$$\nu(p) = \nu(p_0, a_0)\nu(p_1, a_1)\ldots\nu(p_{n-1}, a_{n-1})$$

denote the *output* of the path $p$, with $\nu(p) = \lambda$ if $|p| = 0$ (i.e., if $p = (p_0)$ for some $p_0 \in Q$). Define a path $c = (p_0, \ldots, p_n)$ to be a *cycle* in $T$ if $|c| > 0$ and $p_0 = p_n$. Given a cycle $c$, we say that $c$ is a *$\lambda$-cycle* if $\nu(c) = \lambda$.

Let $T = (Q, \delta, \nu, q_0)$ be a FST, and let $s, f \in Q$. If two unequal paths $p = (s, \ldots, f)$ and $q = (s, \ldots, f)$ from $s$ to $f$ satisfy $\nu(p) = \nu(q)$, we call the pair $(p, q)$ a *bad pair* (for $(s, f)$). The following property of FST's is well-known [9, 10].

**Lemma 3.5.** *A FST is IL if and only if it contains no bad pairs.*

Given a path

$$p = (p_1, a_1, \ldots, p_{n-1}, a_{n-1}, p_n),$$

define a *(proper) 1-step subpath*

$$p' = (p'_1, a'_1, \ldots, p'_{m-1}, a'_{m-1}, p'_m)$$

of $p$, written $p' \prec_1 p$, to be a path satisfying $m < n$ and one of the following conditions:

1. $p'$ is a *proper prefix* of $p$: for all $1 \leq i \leq m$, $p'_i = p_i$, and $a'_i = a_i$ when $i < m$.
2. $p'$ is a *proper suffix* of $p$: for all $1 \leq i \leq m$, $p'_i = p_{i+n-m}$, and $a'_i = a_{i+n-m}$ when $i < m$.
3. $p'$ is a *cycle-reduced subpath* of $p$. This means that there exists a cycle $c$ in $p$ such that removing $c$ from $p$ results in $p'$. For example, if $p$ has a cycle as follows:

$$p = (p_1, a_1, \ldots, \underbrace{p_i, a_i, \ldots, p_j, a_j, p_i}_{\text{cycle}}, b_i, \ldots, p_n),$$

then by removing this cycle, $p$ gives rise to the 1-step subpath

$$p' = (p_1, a_1, \ldots, p_i, b_i, \ldots, p_n).$$

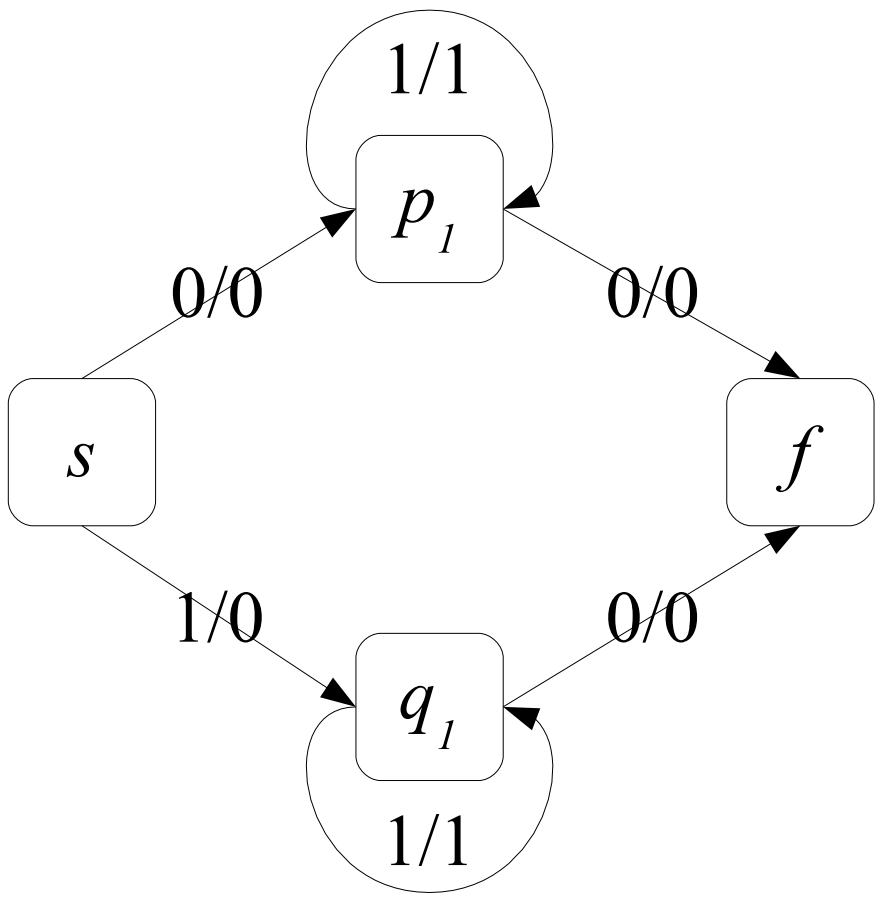

(a) FST with a non-simple bad pair due to cycles

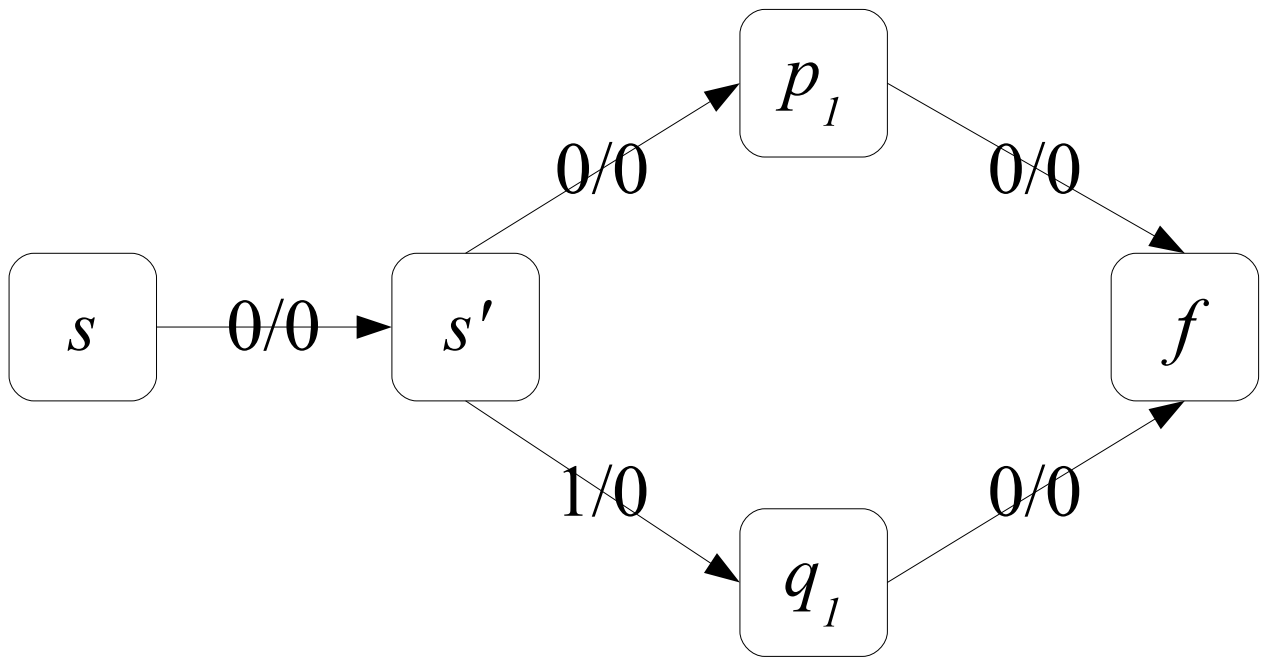

(b) FST with a non-simple bad pair due to overlapping prefixes

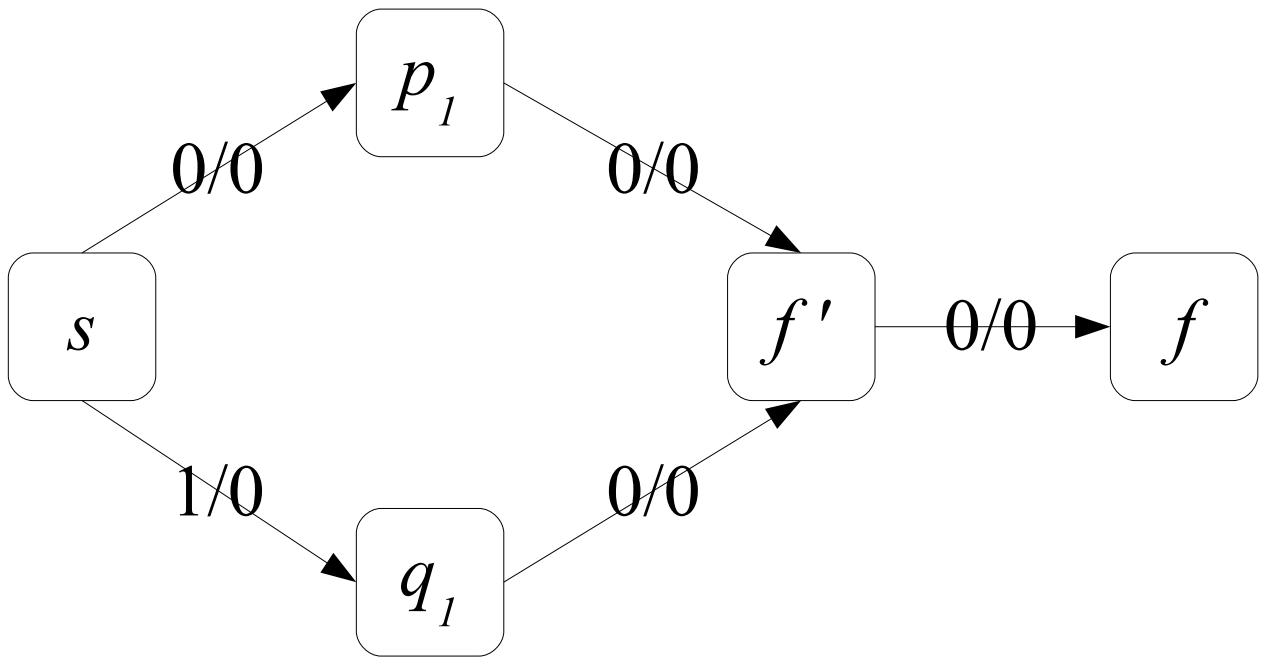

(c) FST with a non-simple bad pair due to overlapping prefixes

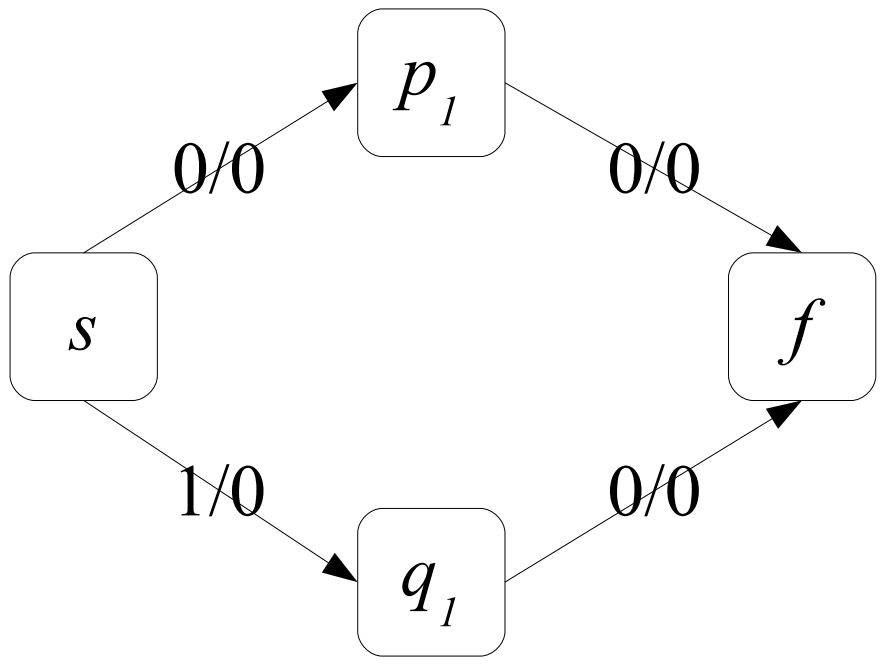

(d) FST with only a simple bad pair

**Figure 3.1:** Examples of three non-simple bad pairs and a simple bad pair. Figure 3.1(a) shows an FST in which the only simple bad pair for $(s, f)$ is the pair of paths $p = (s, 0, p_1, 0, f)$ and $q = (s, 1, q_1, 0, f)$. Other bad sets may be constructed from $p$ and $q$ by adding cycles (e.g. $p' = (s, 0, p_1, 1, p_1, 1, p_1, 0, f)$ and $q' = (s, 1, q_1, 1, q_1, 1, q_1, 0, f)$), but since these can be changed into the bad pair $(p, q)$ by removing the cycles, they are not simple bad pairs. In Figure 3.1(b), there is no simple bad pair for $(s, f)$, although there is a bad pair consisting of the paths $p = (s, 0, s', 0, p_1, 0, f)$ and $q = (s, 0, s', 1, q_1, 0, f)$, which, by removing the prefix $(s, 0)$ from $p$ and $q$, forms a simple bad pair for $(s', f)$. Similarly, in Figure 3.1(c), there is a bad pair, but no simple bad pair, for $(s, f)$, although there is a simple bad pair for $(s, f')$. Finally, in Figure 3.1(d), the only bad pair for $(s, f)$, which is the pair of paths $p = (s, 0, p_1, 0, f)$ and $q = (s, 1, q_1, 0, f)$, is also simple.

Let $\preceq = \prec_1^*$ denote the reflexive, transitive closure of $\prec_1$. We say $p'$ is a *subpath* of $p$ if $p' \preceq p$. We say $p$ is a *proper subpath* of $q$ if $p \preceq q$ and $p \neq q$. Let $s, f \in Q$, and let $(p, q)$ be a bad pair for $(s, f)$. We say $(p, q)$ is a *simple bad pair* for $(s, f)$ (a.k.a., $(p, q)$ is *simple*) if, for all $p' \preceq p$ and $q' \preceq q$, $(p', q')$ is a bad pair if and only if $(p', q') = (p, q)$.

Note that, if a $\lambda$-cycle is removed from a path, then the subpath's output is the same as that of the path, leading to the following observation.

**Observation 3.6.** *If $(p, q)$ is a simple bad pair, then neither $p$ nor $q$ contains a $\lambda$-cycle.*

Intuitively, the paths of a simple bad pair cannot be shrunken through removal of prefixes, suffixes, or cycles, while remaining a bad pair. See Figure 3.1 for an example of three types of bad pairs that are not simple, and one bad pair that is simple. A simple bad pair $(p, q)$ is "canonical" in the sense that the bad pairs formed by its superpaths are bad only because $(p, q)$ is bad, and if $(p, q)$ could be "fixed" somehow, then the bad pairs formed by the superpaths of $p$ and $q$ would be fixed as well. This intuition is reinforced by the following lemma.

**Lemma 3.7.** *A FST is IL if and only if it contains no simple bad pairs.*

*Proof.* Let $T$ be a FST. By Lemma 3.5, $T$ is IL if and only if it contains no bad pairs. Since every simple bad pair is a bad pair, it suffices to show that, if $T$ is not IL, then it contains a simple bad pair.

Assume that $T$ is not IL. Then by Lemma 3.5, $T$ contains a bad pair $(p, q)$. If $(p, q)$ is simple, then the proof is complete. Otherwise, $(p, q)$ is a non-simple bad pair, which means that there exist subpaths $p' \preceq p$ and $q' \preceq q$, at least one of them proper, such that $(p', q')$ is a bad pair. Note that $|p'| + |q'| < |p| + |q|$, since at least one of $p'$ or $q'$ is a proper subpath. Therefore, if $(p', q')$ is not a simple bad pair, we can repeat this process to produce another bad pair $(p'', q'')$ with $|p''| + |q''| < |p'| + |q'|$. However, this sum must be positive for any bad pair. Therefore, the process must eventually terminate with a simple bad pair. $\square$

**Lemma 3.8.** *Let $T = (Q, \delta, \nu, q_0)$ be a FST, let $f \in Q$, and let*

$$X = \{ \ \nu(f', a) \ | \ f' \in Q, a \in \Sigma, \delta(f', a) = f \ \}$$

*be the set of output strings on transition arrows entering $f$. If $X$ is suffix-free and the total number of transition arrows entering $f$ is $|X|$ (i.e., if every such transition arrow has a unique output string), then $f$ is not the final state of any simple bad pair.*

*Proof.* Let $T$, $f$, and $X$ be as in the statement of the lemma. Then by the definition of a simple bad pair, for any simple bad pair $(p, q)$ ending in $f$, the final transition arrows $(f'_p, a_p, f)$ and $(f'_q, a_q, f)$ must be different. Otherwise, $p$ and $q$ could have their last transition removed and remain a bad pair, and $(p, q)$ would not be simple. But since $X$ is suffix-free, $\nu(p) \neq \nu(q)$, so $(p, q)$ cannot be a simple bad pair. $\square$

**Lemma 3.9.** *Every FST has a finite number of simple bad pairs.*

*Proof.* Let $T = (Q, \delta, \nu, q_0)$ be a FST. Let $l = \max_{s \in Q, a \in \Sigma}\{|\nu(s,a)|\}$ be the length of the longest output string on any transition arrow in $T$. Then for any path $p$ in $T$, $|\nu(p)| \leq l|p|$. Note that, if $p$ contains no $\lambda$-cycles, then $|p| \leq |Q||\nu(p)|$. By Observation 3.6, if $(p,q)$ is a simple bad pair, then neither $p$ nor $q$ contains a $\lambda$-cycle. Thus, for any simple bad pair $(p,q)$, since $\nu(p) = \nu(q)$, $|p| \leq |Q||\nu(p)| = |Q||\nu(q)| \leq |q||Q|l$ and likewise, $|q| \leq |p||Q|l$. Therefore, for any $N \in \mathbb{N}$, there are only a finite number of simple bad pairs $(p,q)$ for which $|p| \leq N$ or $|q| \leq N$. We will complete the proof by showing that any bad pair $(p,q)$ such that $|p|, |q| > l|Q|^3|\Sigma|^l$ cannot be simple.

Let $(p,q)$ be a bad pair such that $|p| > l|Q|^3|\Sigma|^l$ and $|q| > l|Q|^3|\Sigma|^l$. If $p \upharpoonright 1 = q \upharpoonright 1$, then $(p,q)$ is not simple, so assume that $p \upharpoonright 1 \neq q \upharpoonright 1$. We proceed through the paths $p$ and $q$ in stages in an attempt to "approximately synchronize" their outputs. For each stage $n \in \mathbb{N}$, define the positions $i_n, j_n \in \mathbb{N}$ (with $i_0 < i_1 < \ldots$ and $j_0 < j_1 < \ldots$) recursively as follows. $i_0 = j_0 = 0$. For all $n \in \mathbb{N}$, let $i_{n+1}$ be the smallest integer such that $\nu\left(q \upharpoonright j_n\right) \sqsubset \nu\left(p \upharpoonright i_{n+1}\right)$, and let $j_{n+1}$ be the smallest integer such that $\nu\left(p \upharpoonright i_{n+1}\right) \sqsubseteq \nu\left(q \upharpoonright j_{n+1}\right)$. $i_n$ and $j_n$ ensure that the output of path $q$ at stage $n$ is at least as long as the output of path $p$ at stage $n$, but no longer than is necessary to ensure that this holds, and the output of path $p$ at the stage $n+1$ is just long enough to extend the output of path $q$ at stage $n$.

For all stages $n \geq 0$, $0 < |\nu(p \upharpoonright i_{n+1})| - |\nu(p \upharpoonright i_n)| \leq l$, $0 < |\nu(q \upharpoonright j_{n+1})| - |\nu(q \upharpoonright j_n)| \leq l$, and $0 \leq |\nu(q \upharpoonright j_n)| - |\nu(p \upharpoonright i_n)| < l$. In other words, the length of the output between successive stages in either path grows by at most $l$, and, in any stage, the amount by which the length of $q$'s output exceeds the length of $p$'s output at that stage is less than $l$. These bounds follow from the definition of $i_n$ and $j_n$.

For all $n \geq 0$, let $p_n$ be the final state of $p \upharpoonright i_n$, let $q_n$ be the final state of $q \upharpoonright j_n$, and let $u_n \in \Sigma^{<l}$ be the string such that $\nu(q \upharpoonright j_n) = \nu(p \upharpoonright i_n)u_n$, the "extra extension" of the output of path $q$ at stage $n$. Note that each triple of the form $(p_n, q_n, u_n)$ is an element of the *finite* set $Q \times Q \times \Sigma^{<l}$, of cardinality less than $|Q|^2|\Sigma|^l$. As noted earlier, since there are no $\lambda$-cycles in $p$ or $q$, the length of any path is at most $|Q|$ times the length of its output. Because $|p|, |q| > l|Q|^3|\Sigma|^l$, it follows that $|\nu(p)|, |\nu(q)| > l|Q|^2|\Sigma|^l$. Since the length of either output increases by at most $l$ with each stage, there are more than $|Q|^2|\Sigma|^l$ stages. By the pigeonhole principle, at least one triple $(p_i, q_i, u_i) \in Q \times Q \times \Sigma^{<l}$ must appear twice in the stage-by-stage enumeration $(p_0, q_0, u_0), (p_1, q_1, u_1), \ldots$.

Let $0 < i < j$ represent two different stages such that $(p_i, q_i, u_i) = (p_j, q_j, u_j)$. Then $c_p = (p_i, \ldots, p_j)$ and $c_q = (q_i, \ldots, q_j)$ each represent a cycle in $p$ and $q$, respectively, of the same output length. While these cycles do not have the same output, $\nu(c_p)$ is a "shifted" version of $\nu(c_q)$: $\nu(c_p)u_i = u_i\nu(c_q)$. Therefore, removing $c_p$ from $p$ and $c_q$ from $q$ will create two different subpaths $p' \preceq p$ and $q' \preceq q$ such that $\nu(p') = \nu(q')$. Since $i > 0$, $p' \upharpoonright 1 = p \upharpoonright 1 \neq q \upharpoonright 1 = q' \upharpoonright 1$, therefore $p' \neq q'$. Because $\nu(p') = \nu(q')$, $(p', q')$ is a bad pair, whence $(p,q)$ is not simple. □

The following theorem is the main theorem of this paper. It establishes that, unlike the trivial case of compression, up to a constant change in the size of the FST's, lossy FST's cannot achieve better *decompression* than ILFST's.

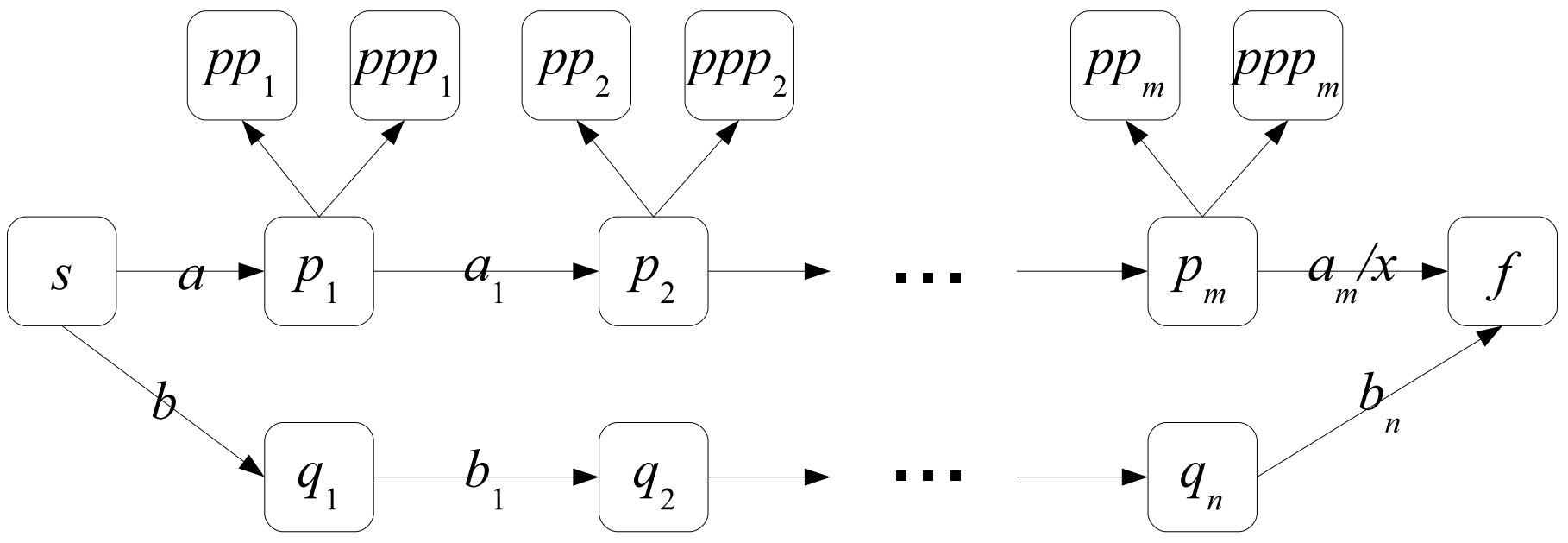

(a) $T$ before the procedure

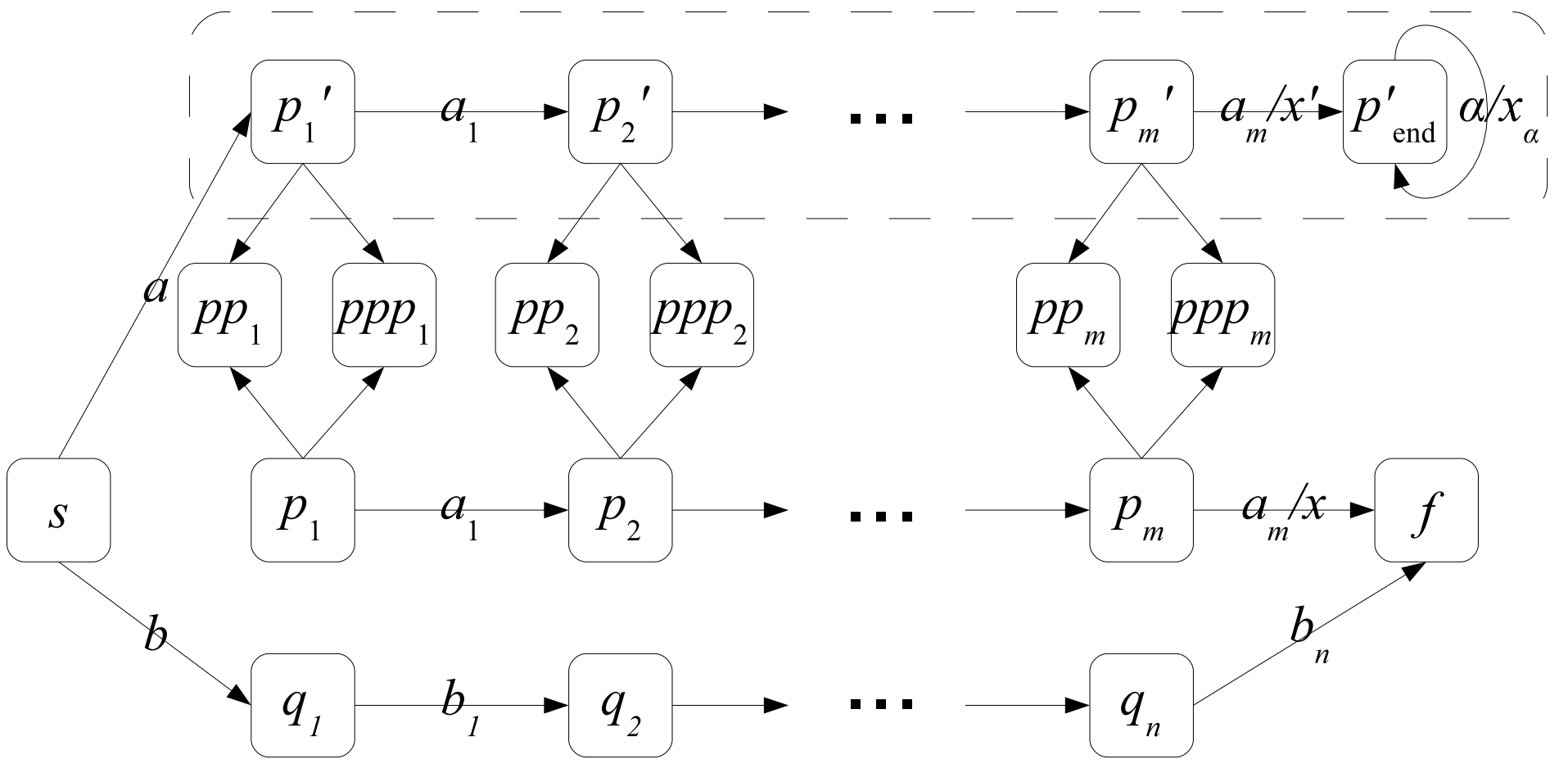

(b) $T$ after the procedure

**Figure 3.2:** Part of a lossy FST $T$ before and after the procedure to eliminate one simple bad pair. Transition arrows are labeled incompletely for readability, and most output strings are not shown; the full formal description of the transformation is given in the text. Figure 3.2(a) illustrates that there exist two different paths from some vertex $s$ to some vertex $f$ that produce the same output. Here, $m \geq n$. States $pp_1$ and $ppp_1$ are other successor states of $p_1$ (besides $p_2$), and likewise with the states $pp_2, ppp_2$, etc. Figure 3.2(b) shows that, to prevent both paths from reaching $f$ with the same output, the upper path is "cloned" by creating clones of the states (indicated with a prime) comprising the upper path, and sending $T$ along this new path instead, if the symbol $a$ is read. The new states are shown surrounded by dashed lines. The new path completely duplicates the behavior of the old path (because each cloned state also clones the outgoing transition arrows, including the output strings), unless the second-to-last state of the path, $p'_m$, is reached. In this case, instead of going to state $f$, $T$ goes to state $p'_{\text{end}}$ (and outputs a string $x'$ possibly different from $x$), all of whose transition arrows self-loop. Since the set $X = \{ \ x_\alpha \ \mid \alpha \in \Sigma \ \} \cup \{x'\}$ of in transition arrows to $p'_{\text{end}}$ is suffix-free, $p'_{\text{end}}$ cannot be the end state of a simple bad pair. Intuitively, the states $p'_1, \ldots, p'_m$ behave exactly like $p_1, \ldots, p_m$, but they remember that they were reached via $s$, and they prevent $T$ from entering state $f$ at the end of the path and thereby losing information.

**Theorem 3.10.** *For all $k \in \mathbb{N}$, there exists $k' \in \mathbb{N}$ such that, for all $x \in \Sigma^*$,*

$$\mathrm{D}_{\mathrm{ILFS}}^{k'}(x) \leq \mathrm{D}_{\mathrm{FS}}^{k}(x).$$

*Proof.* Let $k \in \mathbb{N}$, and let $T = (Q, \delta, \nu, q_0) \in \mathrm{FST}^{\leq k}$ be a lossy FST. We construct an ILFST $T'$ such that, for every minimal program $\pi \in \Sigma^*$ for $T$, there is a program $\pi' \in \Sigma^{|\pi|}$ such that $T(\pi) = T'(\pi')$. In other words, for all $x \in \Sigma^*$, the shortest program for $T'$ that outputs $x$ is no larger than the shortest program for $T$ that outputs $x$. Since $|\mathrm{FST}^{\leq k}|$ is finite, this establishes the theorem with $k' = \max_{T \in \mathrm{FST}^{\leq k}} |T'|$.

We proceed as follows. By Lemma 3.7, if $T$ is not IL, then it has one or more simple bad pairs. By Lemma 3.9, it has a finite number of these. The construction of $T'$ from $T$ will simply eliminate these simple bad pairs one by one, while ensuring that, for each $n \in \mathbb{N}$ such that there is a minimal program of length $n$ for a string $x$, at least one program of length $n$ for $x$ remains. Of course, even though there are a finite number of simple bad pairs, it may be the case that the procedure to eliminate one simple bad pair introduces others. At the conclusion of the proof we demonstrate how to account for this.

Let $s, f \in Q$, $x \in \Sigma^*$, and let $(p, q)$ be a simple bad pair for $(s, f)$ with output $x$. Figure 3.2 shows the part of $T$ relevant to the simple bad pair $(p, q)$, and it illustrates the procedure to eliminate this simple bad pair, which we now describe formally.

Write $p = (s, a, p_1, a_1, p_2, a_2, \ldots, p_m, a_m, f)$ and $q = (s, b, q_1, b_1, q_2, b_2, \ldots, q_n, b_n, f)$. Since $(p, q)$ is simple, $a \neq b$ and either $p_m \neq q_n$ or $a_m \neq b_n$ (i.e., $p$ and $q$ have different first and last transition arrows).

Assume without loss of generality that $m \geq n$, i.e., that $|p| \geq |q|$. Then, if $m > n$, no minimal program will ever (completely) traverse the path $p$, since any program traversing $p$ can be converted to a smaller program, producing the same output, by traversing $q$ instead. If $m = n$, then it may be the case that a minimal program traverses $p$. However, any program that traverses $p$ can be converted into a program of the *same length* that traverses $q$ instead. Hence, if there is a minimal program that traverses $p$, then there is another minimal program producing the same output that never traverses $p$. We will remove $p$ from $T$ in such a way that $T$'s output will remain unaltered on any program that never traverses $p$.

To remove the bad pair $(p, q)$, we alter $T$'s state set and transition and output functions in the following way. Add the states $p'_1, p'_2, \ldots, p'_m, p'_{\mathrm{end}}$ to $Q$. Note that even if the path $p$ contains cycles and so has fewer than $m$ unique states, we add exactly $m + 1$ unique new states to $Q$ (i.e., we "unroll" any cycles in $p$). Choose a suffix-free set $X \subseteq \Sigma^*$ such that $|X| = |\Sigma| + 1$. Assign to each $\alpha \in \Sigma$ a unique element $x_\alpha \in X$, and let $x' \in X$ denote the remaining element of $X$ not assigned to any $\alpha \in \Sigma$. Alter the transition and output functions from $\delta$ and $\nu$ to $\delta'$ and $\nu'$, respectively, as follows.

1. Let $\delta'(s, a) = p'_1$.
2. For all $1 \leq i < m$, let $\delta'(p'_i, a_i) = p'_{i+1}$ and $\nu'(p'_i, a_i) = \nu(p_i, a_i)$.
3. For all $1 \leq i \leq m$ and $\alpha \in \Sigma - \{a_i\}$, let $\delta'(p'_i, \alpha) = \delta(p_i, \alpha)$ and $\nu'(p'_i, \alpha) = \nu(p_i, \alpha)$.

4. Let $\delta'(p'_m, a_m) = p'_{\text{end}}$ and $\nu'(p'_m, a_m) = x'$.

5. For all $\alpha \in \Sigma$, let $\delta'(p'_{\text{end}}, \alpha) = p'_{\text{end}}$ and $\nu'(p'_{\text{end}}, \alpha) = x_\alpha$.

Let $p' = (s, a, p'_1, a_1, p'_2, a_2, \ldots, p'_m, a_m, p'_{\text{end}})$ denote the new path taken by strings that would have traversed the path $p$.

Let $T\backslash p$ denote the FST obtained by altering $T$ in this way. It is clear that $T(\pi) = T\backslash p(\pi)$ for any program $\pi \in \Sigma^*$ that, when given as input to $T$, never causes $T$ to traverse $p$. Since every output string $x$ of $T$ has at least one minimal program for $T$ that never traverses $p$, this alteration does not increase the complexity of any string relative to $T\backslash p$.

Recall that there are a finite number of simple bad pairs in $T$. We now demonstrate that repeated application of this procedure to $T$ will eventually rid $T$ of all simple bad pairs, even if new simple bad pairs are introduced by the procedure itself. Consider how new simple bad pairs may be introduced by the procedure just described. Since the only existing transition arrow that moved was $(s, a, p_1)$ (to $(s, a, p'_1)$), and since this is the *only* transition arrow into any state of the path $p'$ from outside of $p'$ (see the dashed lines in Figure 3.2), for any new simple bad pair $(r, t)$, either (1) $(r, t)$ must have one of its paths traverse this transition arrow, or (2) $(r, t)$ must originate from one of the "cloned" states $p'_1, \ldots, p'_m$.

(1) Suppose that a new simple bad pair $(r, t)$ has a path (assume it is $r$) that traverses the transition arrow $(s, a, p'_1)$. Then, either (1a) $r$ continues all the way along the path $p'$, (1b) $r$ terminates on $p'_i$ for some $i$, or (1c) $r$ leaves $p'$ before reaching $p'_{\text{end}}$.

   (a) If $r$ continues all the way to $p'_{\text{end}}$, then, because the set of outputs on the transition arrows into $p'_{\text{end}}$ is a suffix-free set, by Lemma 3.8, $p'_{\text{end}}$ cannot be the end state of a simple bad pair, so $(r, t)$ is not a simple bad pair.

   (b) If $r$ terminates on $p'_i$ for some $i$, then note that $p'_i$ has only one in transition arrow, and any singleton set is trivially suffix-free. By Lemma 3.8, $p'_i$ is not the end state of a simple bad pair, so $(r, t)$ is not a simple bad pair.

   (c) If $r$ leaves $p'$ before reaching $p'_{\text{end}}$, then, since the path $p'$ otherwise replicates the behavior of $p$, this will result in a new simple bad pair that simply replaces an old simple bad pair that was destroyed when the transition arrow $(s, a, p_1)$ was removed. Therefore, no paths of *net* new simple bad pairs traverse the transition arrow $(s, a, p'_1)$.

(2) Suppose that a new simple bad pair $(r', t')$, where $r' = (p'_i, a_r, \ldots)$ and $t' = (p'_i, a_t, \ldots)$, originates from one of the cloned states $p'_1, \ldots, p'_m$. (1a) and (1b) tell us that no simple bad pair can end in any state along $p'$. Thus, $(r', t')$ is "equivalent" to an existing simple bad pair $(r, t)$, where $r = (p_i, a_r \ldots)$ and $t = (p_i, a_t \ldots)$, in the sense that their paths traverse the same states, except for the fact that one of $r'$ (resp. $t'$) has an initial segment that traverses the cloned states $p'_i, p'_{i+1}, \ldots$ instead of $p_i, p_{i+1}, \ldots$ for a time before leaving $p'$ and "rejoining" with $r$ (resp. $t$). Given two simple bad pairs, place them in the same equivalence class if they were initially the same simple bad pair, but are now different because their first state was cloned. While $(r, t)$ and $(r', t')$ are not the same simple bad pair, they are in the same equivalence class. When the simple bad pair $(r, t)$ is fixed by the procedure, the transition arrow $(p_i, a_r, r_1)$ will be redirected to $(p_i, a_r, r''_1)$, where $r''_1$ is the first state in a new path $r''$ introduced into $T$ ($r''_1$ plays the same role as $p'_1$ did in the bad pair $(p, q)$). To simultaneously fix $(r', t')$, we redirect the transition arrow $(p'_i, a_r, r'_1)$ to $(p'_i, a_r, r''_i)$ as well; in other words, if $T$ is in state $p'_i$ and reads $a_r$, send $T$ along the same new path $r''$ to which it would be redirected if it were in state $p_i$. Since $r''$ ensures the path $r$ does not lose information, $r''$ will do so for $r'$ as well. Of course, it is possible for the procedure to make a clone of a clone, which would admit more than 2 simple bad pairs to the same equivalence class; this case is handled in the obvious way, where all of the longer paths of each simple bad pair in the class would be redirected to the same newly created path in one step.

Since we have altered the procedure to fix all simple bad pairs in an equivalence class in one step, it is clear that the number of equivalence classes will decrease by one each time the procedure is applied. Since each equivalence class described in (2) will correspond to one simple bad pair that was present in the initial FST $T$, by iteratively applying this procedure to each equivalence class of simple bad pairs, all simple bad pairs will be eliminated in a finite number of steps. □

Theorem 3.10 implies a new characterization of the finite-state dimension of individual sequences in terms of decompression by (possibly lossy) finite-state transducers.

**Theorem 3.11.** *For all $S \in \Sigma^\infty$,*

$$\dim_{\mathrm{FS}}(S) = \lim_{k\to\infty} \liminf_{n\to\infty} \frac{\mathrm{D}^k_{\mathrm{FS}}(S \upharpoonright n)}{n},$$

*and*

$$\mathrm{Dim}_{\mathrm{FS}}(S) = \lim_{k\to\infty} \limsup_{n\to\infty} \frac{\mathrm{D}^k_{\mathrm{FS}}(S \upharpoonright n)}{n}.$$

*Proof.* Since every ILFST is a FST, for all $k \in \mathbb{N}$ and $x \in \Sigma^*$, $\mathrm{D}^k_{\mathrm{ILFS}}(x) \geq \mathrm{D}^k_{\mathrm{FS}}(x)$. The theorem follows by Theorem 3.10 and Lemma 3.4. □

Finally, we note that an analog of the definition of finite-state dimension given in §2 holds for lossy decompressors as well. Given a sequence $R \in \Sigma^\infty$ and a FST $T$, define $T(R)$ to be the *output* of $T$ on $R$, the shortest element $S \in \Sigma^\infty \cup \Sigma^*$ such that, for all $n \in \mathbb{N}$, $T(R \upharpoonright n) \sqsubseteq S$.

**Theorem 3.12.** *For all $S \in \Sigma^\infty$,*

$$\dim_{\mathrm{FS}}(S) = \inf_{\substack{T \in \mathrm{FST}, R \in \Sigma^\infty \\ T(R)=S}} \liminf_{m\to\infty} \frac{m}{|T(R \upharpoonright m)|},$$

*and*

$$\mathrm{Dim}_{\mathrm{FS}}(S) = \inf_{\substack{T \in \mathrm{FST}, R \in \Sigma^\infty \\ T(R)=S}} \limsup_{m\to\infty} \frac{m}{|T(R \upharpoonright m)|}$$

*Proof.* We show the result for $\dim_{\mathrm{FS}}$; the proof for $\mathrm{Dim}_{\mathrm{FS}}$ is analogous. By Theorem 3.3,

$$\begin{aligned}
\dim_{\mathrm{FS}}(S) &\triangleq \inf_{T \in \mathrm{ILFST}} \liminf_{n\to\infty} \frac{|T(S \upharpoonright n)|}{n} \\
&= \inf_{\substack{T \in \mathrm{ILFST}, R \in \Sigma^\infty \\ T(R)=S}} \liminf_{m\to\infty} \frac{m}{|T(R \upharpoonright m)|} \\
&\geq \inf_{\substack{T \in \mathrm{FST}, R \in \Sigma^\infty \\ T(R)=S}} \liminf_{m\to\infty} \frac{m}{|T(R \upharpoonright m)|} \\
&\geq \lim_{k\to\infty} \liminf_{n\to\infty} \frac{\mathrm{D}^k_{\mathrm{FS}}(S \upharpoonright n)}{n} \\
&= \dim_{\mathrm{FS}}(S),
\end{aligned}$$

because allowing the FST's to be lossy, and allowing a different FST of length $\leq k$ for each prefix of $S$, cannot increase the complexity of $S$. ☐

**Acknowledgments.** We thank Jack Lutz for pointing out [10] and for advice in preparing this paper, and Elvira Mayordomo for suggesting the addition of Theorem 3.12. We thank the organizers of the Second Computability in Europe Conference, where part of this research was completed, and the Association for Symbolic Logic, which funded part of the first author's travel to that meeting.